\documentclass[10pt]{llncs}
\usepackage{graphics,amssymb,amsmath,epsfig}
\usepackage{longtable}
\makeatletter

\def\vp{{\rm VP}}
\def\vnp{{\rm VNP}}
\def\vdet{{\rm VDET}}
\def\per{{\rm per}}
\def\ham{{\rm ham}}
\def\p{{\rm P}}
\def\np{{\rm NP}}
\def\sharpp{{\rm \sharp P}}

\makeatletter

\begin{document}

\title{On the expressive power of planar perfect matching\\
and permanents of bounded treewidth matrices
}

\author{Uffe Flarup
\inst{1},
Pascal Koiran
\inst{2}
and
Laurent Lyaudet
\inst{2}
}
\institute{Department of Mathematics and Computer Science\\
Syddansk Universitet,
Campusvej 55, 5230 Odense M, Denmark\\
e--mail: \{flarup\}@imada.sdu.dk; fax: +45 65 93 26 91\\
\and Laboratoire de l'Informatique du Parall\'elisme\footnote{UMR 5668 ENS Lyon,
  CNRS, UCBL, INRIA. Research Report RR2007-20}\\
Ecole Normale Sup\'erieure de Lyon,
46, all\'ee d'Italie, 69364 Lyon Cedex 07, France\\
e--mail: \{pascal.koiran,laurent.lyaudet\}@ens-lyon.fr; fax: +33 4 72 72 80 80
}

\maketitle

\begin{abstract}
Valiant introduced some 25 years ago an algebraic model of computation
along with the complexity classes $\vp$ and $\vnp$, 
which can be viewed as  analogues of the
classical classes $\p$ and $\np$.
They are defined using non-uniform sequences of arithmetic circuits and
provides a framework to study the complexity for sequences of polynomials.
Prominent examples of difficult (that is, $\vnp$-complete) problems in
this model includes the permanent and hamiltonian polynomials.

While the permanent and hamiltonian polynomials in general are
difficult to evaluate, there have been research on which special cases
of these polynomials admits efficient evaluation. For instance, 
Barvinok has
shown that if the underlying matrix has bounded rank, both the permanent and
the hamiltonian polynomials can be evaluated in polynomial time,
and thus are in $\vp$.
Courcelle, Makowsky and Rotics have shown that for matrices of bounded 
treewidth several difficult problems (including evaluating
the permanent and hamiltonian polynomials) can be solved efficiently.
An earlier result of this flavour is Kasteleyn's theorem which
states that the sum of weights of perfect matchings of a planar graph
can be computed in polynomial time, and thus is in $\vp$ also.
For general graphs this problem is $\vnp$-complete.

In this paper we investigate the expressive power of the above
results. We show that the permanent and hamiltonian polynomials for matrices
of bounded treewidth both are equivalent to arithmetic formulas.
Also, arithmetic weakly skew circuits are shown to be equivalent to the sum
of weights of perfect matchings of planar graphs.


\end{abstract}

\section{Introduction}

Our focus  in this paper is on easy special cases of otherwise difficult to
evaluate polynomials, and their connection to various classes of
arithmetic circuits.
In particular we consider the permanent and
hamiltonian polynomials for matrices of bounded treewidth,
and sum of weights of perfect
matchings for planar graphs.
It is a widely believed conjecture that the permanent is hard to evaluate.
Indeed, in Valiant's framework~\cite{Valiant79,Valiant82}
 the permanent is complete for the class \vnp.
This is an algebraic analogue of his $\sharpp$-completeness result for the
permanent~\cite{Valiant79a}.
For a book-length treatment of Valiant's algebraic complexity theory
one may consult~\cite{Burg}.
The same results ($\sharpp$-completeness in the boolean framework,
and \vnp-completeness in the algebraic framework) 
also apply to the hamiltonian polynomial.
The sum of weights of perfect matchings in an (undirected) graph $G$
is yet another example of a presumably hard to compute polynomial since
it reduces to the permanent when $G$ is bipartite.
However, all three polynomials are known to be easy to evaluate 
in special cases. In particular, the permanent and hamiltonian polynomials
can be evaluated in a polynomial number of arithmetic operations for matrices 
of bounded treewidth~\cite{CMR}.
An earlier result of this flavour is Kasteleyn's theorem \cite{Ka} which
states that the sum of weights of perfect matchings of a planar graph
can be computed in a polynomial number of arithmetic operations.
One can try to turn these three efficient algorithms into 
general-purpose evaluation algorithms by means of reductions
(this is the approach followed by Valiant in~\cite{Va04}, 
where he exhibits polynomial
time algorithms for several problems which previously only had exponential time
algorithms, by means of holographic reductions to perfect
matchings of planar graphs).
For instance, in order to evaluate a polynomial $P$ one can try to construct
a matrix of bounded treewidth $A$ such that:
\begin{itemize}
\item[(i)] The entries of $A$ are variables of $P$, or constants.
\item[(ii)] The permanent of $A$ is equal to $P$.
\end{itemize}
The same approach can be tried for the hamiltonian 
and the sum of weights of perfect matchings in a planar graph.
The goal of this paper is to assess the power of these polynomial
evaluation methods.
It turns out that the three  methods  are all universal - 
that is, every polynomial can be expressed 
as the sum of weights of perfect matchings in a planar graph,
and as a permanent and hamiltonian  of matrices of bounded treewidth.
>From a complexity-theoretic point of view, 
these methods are no longer equivalent.
Our main findings are that:
\begin{itemize}
\item The permanents and hamiltonians of matrices of polynomial size 
and bounded treewidth
have the same expressive power, namely, the power of polynomial 
size arithmetic formulas. This is established in Theorem~\ref{treewidthchar}.
\item The sum of weights of perfect matchings in polynomial size 
planar graphs has at least the same power as the above two representations,
and in fact it is more powerful under a widely believed conjecture.
Indeed, this representation has the same power as polynomial size (weakly)
skew arithmetic circuits. This is established in Theorem~\ref{planarchar}.
We recall that (weakly)
skew arithmetic circuits capture the complexity 
of computing the determinant~\cite{To}.
It is widely believed that the determinant cannot be expressed by polynomial
size arithmetic formulas.
\end{itemize}
Our three methods therefore capture (presumably proper) subsets of
the class \vp\ of easy to compute polynomial families.
By contrast, if we drop the bounded treewidth or planarity assumptions,
the class \vnp\ is captured in all three cases.

Various notions of graph ``width'' have been defined in the litterature besides
treewidth (there is for instance pathwidth, cliquewidth, rankwidth...).
They should be worth studying from the point of view of their expressive power.
Also, Barvinok~\cite{Ba} has
shown that if the underlying matrix has bounded rank, both the permanent and
the hamiltonian polynomials can be evaluated in a polynomial number of
arithmetic operations.
A proper study of the expressive power of permanents and hamiltonians of 
bounded rank along the same line as in this paper remains to be done.

\newpage


\section{Definitions}

\subsection{Arithmetic circuits}

\begin{definition}
An arithmetic circuit is a finite, acyclic, directed graph. Vertices have
indegree 0 or 2, where those with indegree 0 are referred to as {\em inputs}.
A single vertex must have outdegree 0, and is referred to as {\em output}.
Each vertex of indegree 2 must be labeled by either $+$ or $\times$, thus
representing computation. Vertices are commonly referred to as {\em gates}
and edges as {\em arrows}.
\end{definition}

By interpreting the input gates either as constants or variables it is
easy to prove by induction that each arithmetic circuit naturally
represents a polynomial.

In this paper various subclasses of arithmetic circuits will be considered:
For {\em weakly skew} circuits we have the restriction that for every
multiplication gate, at least one of the incoming arrows is from a subcircuit
whose only connection to the rest of the circuit is through this incoming
arrow.
For {\em skew} circuits we have the restriction that for every
multiplication gate, at least one of incoming arrows is from an
input gate.
For {\em formulas} all gates
(except output) have outdegree 1. Thus, reuse of partial
results is not allowed.

For a detailed description of various subclasses of arithmetic circuits,
along with examples, we refer to \cite{MP}.

\begin{definition}
The {\em size} of a circuit is the total number of {\em gates} in the circuit.
The {\em depth} of a circuit is the length of the longest path from an
input gate to the output gate.
\end{definition}

A family $(f_n)$ belongs to the complexity class \vp\ if $f_n$ can be computed by a circuit $C_n$ of size polynomial in $n$, and if moreover 
the degree of $f_n$ is bounded by a polynomial function of~$n$.

\subsection{Treewidth}

Treewidth for undirected graphs is most commonly defined as follows:

\begin{definition}
Let $G = \langle V,E \rangle$ be a graph. A $k$-tree-decomposition of $G$ is:
\begin{itemize}
\item[(i)] A tree $T = \langle V_T, E_T \rangle$.
\item[(ii)] For each $t \in V_T$ a subset $X_t \subseteq V$
of size at most $k + 1$.
\item[(iii)] For each edge $(u,v) \in E$ there is a $t \in V_T$ such that
$\lbrace u,v \rbrace \subseteq X_t$.
\item[(iv)] For each vertex $v \in V$ the set $\lbrace t \in V_T |
v \in X_T \rbrace$ forms a (connected) subtree of $T$.
\end{itemize}
The treewidth of $G$ is then the smallest $k$ such that there exists a
$k$-tree-decomposition for $G$.
\end{definition}

There is an equivalent definition of treewidth in terms of certain graph grammars called HR algebras~\cite{Co}:
\begin{definition}
A graph $G$ has a $k$-tree-decomposition iff there exist a set of source
labels of cardinality $k+1$ such that $G$ can be constructed using
a finite number of the following operations:
\begin{itemize}
\item[(i)] $ver_a$, $loop_a$, $edge_{ab}$ (basic constructs: create a single
vertex with label $a$, a single vertex with label $a$ and a looping edge, two
vertices labeled $a$ and $b$ connected by an edge)
\item[(ii)] $ren_{a \leftrightarrow b} (G)$ (rename all labels $a$ as labels
$b$ and rename all labels $b$ as labels $a$)
\item[(iii)] $forg_a (G)$ (forget all labels $a$)
\item[(iv)] $G_1 \; // \; G_2$ (composition of graphs: any two
vertices with the same label are identified as a single vertex)
\end{itemize}
\end{definition}

\begin{example}
Cycles are known to have treewidth 2. Here we show that they
have treewidth at most 2 by constructing
$G$, a cycle of length $l \geq 3$, using $\lbrace a,b,c \rbrace$ as the
set of source labels.
First we construct $G_1$ by the operation $edge_{ab}$.
For $1 < i < l$ we construct $G_i$ by operations
$forg_c(ren_{b \leftrightarrow c}(G_{i-1} \; // \; edge_{bc})$.
Finally $G$ is then constructed by the operation $G_{l-1} \; // \; edge_{ab}$.
\end{example}

The treewidth of a directed graph is defined as the treewidth of the underlying
undirected graph.
The treewidth of an $(n \times n)$ matrix $M = (m_{i,j})$ is defined as
the treewidth of the directed graph $G_M = \langle V_M,E_M,w \rangle$ where
$V_M = \lbrace 1, \ldots , n \rbrace$, $(i,j) \in E_M$ iff
$m_{i,j} \neq 0$, and $w(i,j) = m_{i,j}$.
Notice that $G_M$ can have loops. Loops do not affect
the treewidth of $G_M$ but are important for the characterization of the
permanent and hamiltonian polynomials.

\subsection{Permanent and hamiltonian polynomials}

In this paper we take a graph theoretic approach to deal with permanent
and hamiltonian polynomials. The reason for this being that a natural way
to define the treewidth of a matrix, is by the treewidth of the underlying
graph, see also e.g. \cite{MM}.

\begin{definition}
A {\em cycle cover}
of a directed graph is a subset of the edges, such that
these edges form disjoint, directed cycles (loops are allowed).
Furthermore, each vertex in the
graph must be in one 
(and only one)
of these cycles. The {\em weight} of a cycle cover
is the {\em product} of weights of all participating edges.
\end{definition}

\begin{definition} \label{permdef}
The {\em permanent} of an $(n \times n)$ matrix $M = (m_{i,j})$ is the 
sum of weights of all cycle covers of $G_M$.
\end{definition}
The permanent of $M$ can also be defined by the formula
$$\per(M)=\sum_{\sigma \in S_n} \prod_{i=1}^n m_{i,\sigma(i)}.$$
The equivalence with Definition~\ref{permdef} is clear since any permutation 
can be written down as a product of disjoint cycles,
and this decomposition is unique.

There is a natural way of representing polynomials by permanents.
Indeed, if the entries of $M$ are variables or constants from some field $K$, 
$f=\per(M)$ is a polynomial with coefficients in $K$ (in Valiant's terminology,
$f$ is a projection of the permanent polynomial).
In the next section we study the power of this representation in the case where
$M$ has bounded treewidth.

The {\em hamiltonian} polynomial
$\ham(M)$ is defined similarly,
except that we only sum over
cycle covers
consisting of a {\em single} cycle (hence the name).

\section{Matrices of bounded treewidth}

In this section we work with directed graphs. All paths and cycles are assumed to be directed, even if this word is omitted.

In \cite{CMR} it is shown that the permanent and hamiltonian polynomials are in
$\vp$ for matrices of bounded treewidth. 
Here we show that both the permanent and hamiltonian polynomials
for matrices of bounded treewidth are equivalent to arithmetic formulas.
This is an improvement on the result of~\cite{CMR} since the set 
of polynomial families
representable by polynomial size arithmetic formulas is a (probably strict)
subset of $\vp$.
\begin{theorem} \label{treewidthchar}
Let $(f_n)$ be a family of polynomials with coefficients in a field $K$.
The three following properties are equivalent:
\begin{itemize}
\item $(f_n)$ can be represented by a family of polynomial size arithmetic 
formulas.
\item There exists a family $(M_n)$ of polynomial size, bounded treewidth
 matrices such that the entries of $M_n$ are constants from $K$ or variables of $f_n$,
and $f_n = \per(M_n)$.
\item There exists a family $(M_n)$ of polynomial size, bounded treewidth
 matrices such that the entries of $M_n$ are constants from $K$ or variables 
of $f_n$,
and $f_n = \ham(M_n)$.
\end{itemize}
\end{theorem}

\begin{remark}
By the $\vnp$-completeness of the hamiltonian, if we drop the 
bounded treewidth assumption on $M_n$ we capture exactly the $\vnp$ families
instead of the families represented by polynomial size arithmetic 
formulas.
The same property holds true for the permanent if the characteristic of $K$ 
is different from 2.
\end{remark}
Theorem~\ref{treewidthchar} follows immediately 
from Theorems~\ref{formulaToPerm}, \ref{formulaToHam}, \ref{permToCircuit}
and~\ref{hamToCircuit}.
\begin{theorem}\label{formulaToPerm}
Every arithmetic formula can be expressed as the permanent
of a matrix of tree\-width at most 2 
and size at most $(n+1) \times (n+1)$
where $n$ is the size of the formula. All entries in the
matrix are either 0, 1, or
variables of the formula.
\end{theorem}
\proof
The first step is to construct a directed graph that is a special case of a
{\em series-parallel} (SP) graph, in which there is a
connection between weights of directed paths and the value computed by
the formula.
The overall idea behind the construction is quite standard,
see e.g. \cite{MP}.
SP graphs in general can between any two adjacent vertices have
multiple directed edges. But we construct an
SP graph in which there is at most one directed edge from any vertex
$u$ to any vertex $v$.
This property will be needed in the second step, in which a connection
between cycle covers
and the permanent of a given matrix will
be established.

SP graphs 
have distinguished {\em source} and {\em sink} vertices,
denoted by $s$ and $t$. By $SW(G)$ we denote the sum of weights of all
directed paths from $s$ to $t$, where the weight of a path is the
{\em product} of weights of participating edges.

Let $\varphi$ be a formula of size $e$. For the first step of the
proof we will by induction over $e$ construct
a weighted, directed SP graph $G$ such that $val(\varphi) = SW(G)$.
For the base case $\varphi = w$ we construct vertices $s$ and $t$ and
connect them by a directed edge from $s$ to $t$ with weight $w$.

Assume $\varphi = \varphi_1 + \varphi_2$ and let $G_i$ be the graph
associated with $\varphi_i$ by the induction hypothesis. Introduce one new
vertex $s$ and let $G$ be
the union of the three graphs $\langle \lbrace s \rbrace
\rangle$, $G_1$ and $G_2$ in which
we identify $t_1$ with $t_2$ and denote it $t$, add an edge of weight
1 from $s$ to $s_1$, and add an edge of weight 1 from $s$ to $s_2$.
By induction hypothesis the resulting graph $G$ 
satisfies $SW(G) = 1 \cdot SW(G_1) + 1 \cdot SW(G_2) = val(\varphi_1) +
val(\varphi_2)$.
Between any two vertices $u$ and $v$ there is at most one directed
edge from $u$ to $v$. We introduced one new vertex, but since $t_1$ was
identified with $t_2$ the number of vertices used equals $|V_1| + |V_2|
\leq size(\varphi_1) + 1 + size(\varphi_2) + 1 = size(\varphi) +1$.

Assume $\varphi = \varphi_1 * \varphi_2$. We construct $G$ by making the
disjoint union of $G_1$ and $G_2$ in which we identify $t_1$ with $s_2$,
identify $s_1$ as $s$ in $G$ and identify $t_2$ as $t$ in $G$. For every
directed path from $s_1$ to $t_1$ in $G_1$ and for every directed path
from $s_2$ to $t_2$ in $G_2$ we can find a directed path from $s$ to $t$
in $G$ of weight equal to the product of the weights of the paths in $G_1$
and $G_2$, and since all $(s,t)$ paths in $G$ are of this type we
get $SW(G) = SW(G_1) \cdot SW(G_2)$. The number of vertices used equals
$|V_1| + |V_2| -1 \leq size(\varphi_1) + size(\varphi_2) + 1
< size(\varphi) +1$.

For the second step of the proof we need to construct a graph
$G'$ such that there is a relation between cycle covers
in $G'$ and
directed paths from $s$ to $t$ in $G$. We construct $G'$ by adding
an edge of weight 1 from $t$ back to $s$, and loops of weight $1$ at
all vertices different from $s$ and $t$.
Now, for every $(s,t)$ path in $G$ we can find a cycle in $G'$ visiting
the corresponding nodes. For nodes in $G'$ not in this cycle, we include
them in a cycle cover
by the loops of weight 1.
Because there is at most one directed edge from any vertex $u$ to any vertex
$v$ in $G'$ we can find a matrix $M$ of size at most $(n+1) \times (n+1)$
such that $G_M = G'$ and
$per(M) = val(\varphi)$.

The graph $G'$ can be constructed using an
HR algebra with only 3 source labels, and thus have treewidth at most 2.
For the base case the operation 
$edge_{ab}$ is sufficient.
For the simulation of addition of formulas the following grammar operations
provide the desired construction:
$ren_{a \leftrightarrow c}
( forg_a (edge_{ac} \; // \;
(loop_a \; // \; G_1)) \; // \;
forg_a (edge_{ac} \; // \; (loop_a \; // \; G_2)))$.
For simulating multiplication of formulas we use
the following grammar operations:
$forg_c(ren_{b \leftrightarrow c}(G_1) \; //$ 
$ren_{a \leftrightarrow c} (loop_a \; // \; G_2))$.
Finally, the last step in obtaining $G'$ is to make a composition
with the graph $edge_{ab}$.
\qed

\begin{figure}
\begin{center}
\includegraphics[scale=0.6]{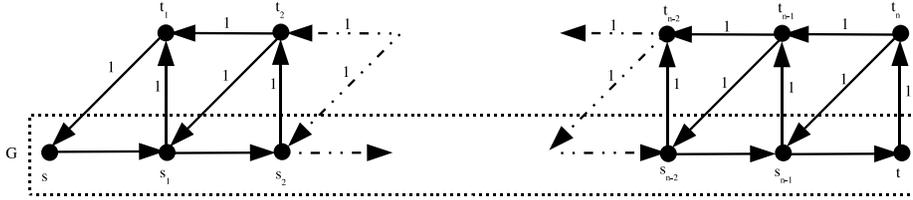}
\caption{Universality of the hamiltonian polynomial}\label{hamUniversal}
\end{center}
\end{figure}

\begin{theorem} \label{formulaToHam}
Every arithmetic formula of size $n$ can be expressed as the hamiltonian
of a matrix of treewidth at most 6 and size at most $(2n+1) \times (2n+1)$.
All entries in the
matrix are either 0, 1, or
variables of the formula.
\end{theorem}
\proof
The first step is to produce the graph $G$ as shown in
Theorem~\ref{formulaToPerm}.
The next step is to show that the proof of universality for the
hamiltonian polynomial in \cite{Ma}
can be done with treewidth at most 6.
Their construction for universality of the hamiltonian polynomial
introduces $|V_G| - 1$ new vertices to $G$ in order to
produce $G'$, along with appropriate directed edges (all of weight 1).
The proof is sketched in Figure~\ref{hamUniversal}.

The additional vertices $t_i$ and edges permit to visit any subset of vertices
of $G$ with a directed path of weight 1 from $t$ to $s$ using all $t_i$'s.
Hence, any path from $s$ to $t$ in $G$ can be followed by a path from $t$
to $s$ to obtain a hamiltonian cycle of same weight.

If one just need to show universality, then it is not important exactly which
one of the vertices $t_i$ that has an edge to a given vertex among $s_i$.
But in order to show
bounded treewidth one carefully need to take into account which one of the
vertices of $t_i$ that has an edge to a particular $s_i$ vertex.
We show such a construction with bounded treewidth, by giving an HR
algebra which can express a graph similar to the one in
Figure~\ref{hamUniversal} using 7 source labels.
\begin{figure}
\begin{center}
\includegraphics[scale=0.55]{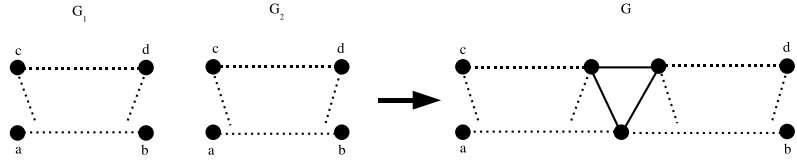}
\caption{Series composition (simulating multiplication)}\label{hamSeries}
\end{center}
\end{figure}

Series composition is done using the following operations (also see
Figure~\ref{hamSeries}):
\begin{align*}
& forg_e(forg_f(forg_g( \\ 
& \quad ren_{d \leftrightarrow f}(ren_{b \leftrightarrow e}(G_1)) \; // \\
& \quad ren_{c \leftrightarrow g}(ren_{a \leftrightarrow e}(G_2)) \; // \\
& \quad edge_{ef} \; // \; edge_{eg} \; // \; edge_{fg} \\
& )))
\end{align*}

\begin{figure}
\begin{center}
\includegraphics[scale=0.55]{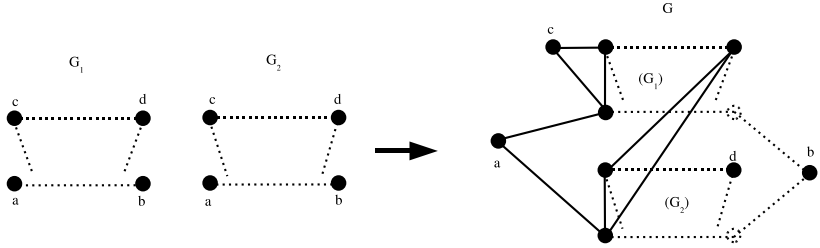}
\caption{Parallel composition (simulating addition)}\label{hamParallel}
\end{center}
\end{figure}

Labels $a, b, c$ and $d$ in Figures~\ref{hamSeries} and~\ref{hamParallel}
plays the roles of $s, t, t_1$ and $t_n$ respectively
in Figure~\ref{hamUniversal}.

The above construction does not take into account, that $G_1$ and/or $G_2$
are graphs generated from the base case. For base cases vertices $c$ and $d$
are replaced by a single vertex. However, it is clear that the above
construction can be modified to work for these simpler cases as well.

For parallel composition an additional vertex was introduced.
It can be done using the following operations (also see
Figure~\ref{hamParallel}):
\begin{align*}
& forg_e(forg_f(forg_g( \\
& \quad ren_{a \leftrightarrow e}(ren_{c \leftrightarrow g}( \\
& \quad \quad forg_a(forg_c(edge_{ag} \; // \; edge_{cg} \; // \; ren_{d
\leftrightarrow f}(edge_{ae} \; // \; edge_{ac} \; // \; G_1))) \; // \\
& \quad \quad forg_a(forg_c(edge_{af} \; // \; edge_{cf} \; // \; edge_{ae} \;
// \; edge_{ac} \; // \; G_2)) \\
& \quad )) \\
& )))
\end{align*}

The final step in the construction, after all series and parallel composition
have been done, is to connect vertices $a$ and $c$ and connect vertices
$b$ and $d$.
\qed

\bigskip

The decision version of the hamiltonian cycle problem for graphs
of bounded cliquewidth is shown to be polynomial time solvable in \cite{EGW}.
Though bounded treewidth implies bounded clique\-width we are mainly interested
in studying the evaluation version. Evaluation of the hamiltonian and
permanent polynomials was shown in \cite{CMR} to be in $\vp$ for matrices
of bounded treewidth. They give efficients algorithms for a much
broader class of problems, but the proof we give here is more direct
and gives a more precise characterization.

By Definition~\ref{permdef}, computing the permanent of a matrix $M$ amounts 
to computing the sum of the weights of all cycle covers of $G_M$.
In our algorithm we need to consider partial covers, 
which are  generalizations of cycle covers.
\begin{definition}
A partial
cover of a directed graph is a union of 
paths and cycles such that every vertex of the graph belongs to at most
one path (and to none of the cycles), or to at most one cycle 
(and to none of the paths). 

The weight of a partial cover is the product of the weights 
of all participating edges. More generally, for any set $S$ of edges the weight
$w(S)$ of $S$ is defined as the product of the weights of the elements of $S$.
\end{definition}
In contrast to cycle covers, for a partial cover there is no requirement 
that all vertices be covered.

The following theorem from~\cite{Bo}
is a standard tool in the design of parallel algorithms for graphs of bounded 
treewidth (see also~\cite{BH} and~\cite{MZ}).

\begin{theorem} \label{balTree}
Let $G = \langle V,E \rangle$ be a graph of treewidth k with n vertices. Then
there exists a tree-decomposition $\langle T, (X_t)_{t \in V_T} \rangle$ of G
of width $3k + 2$ such that $T = \langle V_T,E_T \rangle $ is a binary tree
of depth at most $2 \lceil log_{5 \over 4} (2n) \rceil$.
\end{theorem}

We also need the following standard lemma:

\begin{lemma}\label{circuitToFormula}
Let $\varphi$ be a circuit of depth $d$.
Then there exists a formula of depth $d$ and size $O(2^d)$
representing the same polynomial.
\end{lemma}
\proof
We construct the formula from the circuit by duplicating entire subcircuits
whenever we reuse a gate. The formula constructed in this way also has
height $d$. In the produced formula the number
of gates having distance $j$ to the root is at most
twice the number of gates having distance $j-1$ to the root, so the
formula has at most
$\sum_{i=0}^{d} 2^i = 2 \cdot 2^d - 1$ gates.
\qed

\begin{theorem}\label{permToCircuit}
The permanent of a $n \times n$ matrix $M$ of bounded treewidth $k$ can be
expressed as a formula of size $O(n^{O(1)})$.
\end{theorem}
\proof
We show how to construct a circuit of depth $O(log(n))$, which can then be
expressed as a formula of size $O(n^{O(1)})$ 
using Lemma \ref{circuitToFormula}.
Consider the graph $G=G_M$ and
apply Theorem~\ref{balTree} to obtain a balanced, binary tree-decomposition
$T$ of bounded width $k'$.
For each node $t$ of $T$, we denote by $T_t$ the subtree of $T$ rooted at $t$,
and we denote by $X(T_t)$ the set of vertices of $G$ which belong to $X_u$ for at least one of the nodes $u$ of $T_t$. 
We denote by $G_t$ the subgraph of $G$ induced by
the subset of vertices $X(T_t)$.

Consider a partial cover $C$ of $G_t$. 
Any given edge $(u,v) \in X_t^2$
is either used or unused by $C$. 
Likewise, any given vertex of $X_t$ has
indegree 0 or 1 in $C$, and outdegree 0 or 1. 
We denote by $\lambda_t=I_t(C)$
the list of 
all these data for every edge
$(u,v) \in X_t^2$ and every 
element of $X_t$. By abuse of language, we will say that an edge
in $X_t^2$ 
is used by $\lambda_t$ if it is used by one partial cover satisfying 
$I_t(C)=\lambda_t$ (or equivalently, by all partial cover satisfying 
$I_t(C)=\lambda_t$).

We will compute for each possible list $\lambda_t$ a weight $w_{\lambda_t}$, 
defined as the sum of the weights
of all partial covers $C$ of $G_t$ satisfying the following three properties:
\begin{itemize}
\item[(i)] the two endpoints 
of all paths of $C$ belong to $X_t$;

\item[(ii)] all uncovered vertices belong to $X_t$;
\item[(iii)] $I_t(C)=\lambda_t$.
\end{itemize}
Note that the number of weights to be computed at each node of $T$ 
is bounded by a constant (which depends on $k'$).
When $t$ is the root of $T$ we can easily compute the permanent of $M$ 
from the weights $w_{\lambda_t}$: it is equal to the sums of the $w_{\lambda_t}$
over all $\lambda_t$ which assign indegree 1 and outdegree 1 to all vertices 
of $X_t$.
Also, when $t$ is a leaf of $T$ we can compute the weights 
in a constant number of arithmetic operations since $G_t$ has at most $k'$ 
vertices in this case.
It therefore remains to explain how to compute the weights $w_{\lambda_t}$ 
when $t$ 
is not a leaf. 

Our algorithm for this proceeds in a bottom-up manner: 
we will compute the weights for $t$ from the weights already computed for 
its left child (denoted $l$) and its right child (denoted $r$).
The idea is that we can obtain a partial cover of $G_t$ by taking the union
of a partial cover of $G_l$ and of a partial cover of $G_r$, and adding 
some additional edges. 
Conversely, a partial cover of $G_t$ induces a partial cover of $G_l$ and a 
partial cover of $G_r$.
In order to avoid counting many times the same partial cover, 
we must define the considered partial covers of $G_l$ and $G_r$
to ensure that the partial cover 
of $G_t$ induces a unique partial cover of $G_l$ and a 
unique partial cover of $G_r$.
We will say that $(\lambda_l,\lambda_r)$ is compatible with $\lambda_t$ if and
only if the following holds: 
\begin{itemize} 
\item[-] no edge in $X_t^2$
is used in $\lambda_l$ or $\lambda_r$;
\item[-] for every vertex $x \in X_t$
at most one of $\lambda_t,\lambda_l,\lambda_r$ assigns indegree 1 to $x$;
\item[-] for every vertex $x \in X_t$ at most one of
$\lambda_t,\lambda_l,\lambda_r$ assigns outdegree 1 to $x$;
\item[-] every vertex $x \in X_l\backslash X_t$
has indegree 1 and outdegree 1 in $\lambda_l$;
\item[-] every vertex $x \in X_r\backslash X_t$
has indegree 1 and outdegree 1 in $\lambda_r$.
\end{itemize}
We now have to prove two things. If there is a partial cover $C$ of $G_t$ which satisfies  
the properties (i) and (ii) such that $I_t(C)=\lambda_t$ then it induces a partial cover $C_l$
of $G_l$ and a partial cover $C_r$ of $G_r$ such that $C_l$ and $C_r$ satisfy (i) and (ii),
$I_l(C)=\lambda_l$, $I_r(C)=\lambda_r$, and $(\lambda_l,\lambda_r)$ is compatible with $\lambda_t$.
Conversely, if $(\lambda_l,\lambda_r)$ is compatible with $\lambda_t$, and $C_l$ and $C_r$
are partial covers of $G_l$ and $G_r$ satisfying (i), (ii), $I_l(C)=\lambda_l$, and $I_r(C)=\lambda_r$,
then there exists a unique partial cover $C$ of $G_t$ containing $C_l$ and $C_r$ such that $I_t(C)=\lambda_t$.

Consider a partial cover $C$ of $G_t$ which satisfies  
the properties (i) and (ii) defined above. We can assign to $C$ 
a unique triple $(C_l,C_r,S)$ defined as follows. First, we define $S$ as the
set of edges of $C \cap X_t^2$. 
Then we define $C_l$ as the set of edges of $C$ which have their
two endpoints in $X(T_l)$, and at least one of them outside of $X_t$.
Finally, we define $C_r$ as the set of edges of $C$ which have their
two endpoints in $X(T_r)$, and at least one of them outside of $X_t$.
Note that $w(C)=w(C_l)w(C_r)w(S)$ since 
$(C_l,C_r,S)$ forms a partition of the edges of $C$.
Moreover, $C_l$ is a partial cover of $G_l$ 
and properties (i) and (ii) are satisfied: 
the endpoints of the paths of $C_l$ and the uncovered vertices
of $X(T_l)$  all belong to $X_l\cap X_t$.
Likewise, $C_r$ is a partial cover of $X(T_r)$ 
and properties (i) and (ii) are satisfied.
If $I_l(C)=\lambda_l$ and $I_r(C)=\lambda_r$,
it is clear that $(\lambda_l,\lambda_r)$ is compatible with $\lambda_t$.
Any other partition of $C$ in three parts with one partial cover of $G_l$,
one partial cover of $G_r$, and a subset of edges in $X_t^2$
would have an edge of $X_t^2$
used by $C_l$ or $C_r$. 
Hence $(\lambda_l,\lambda_r)$ would not be compatible with $\lambda_t$.

Suppose now $(\lambda_l,\lambda_r)$ is compatible with $\lambda_t$, and $C_l$ and $C_r$
are partial covers of $G_l$ and $G_r$ satisfying (i), (ii), $I_l(C)=\lambda_l$, and $I_r(C)=\lambda_r$.
We define $S_{\lambda_t}$ as the set of edges of $X_t^2$ which are used by $\lambda_t$. 
It is clear that $S_{\lambda_t}$, $C_l$ and $C_r$ are disjoint.
Consider $C=S_{\lambda_t}\cup C_l \cup C_r$. Since $(\lambda_l,\lambda_r)$ is compatible with $\lambda_t$, 
$C$ is a partial cover satisfying (i) and (ii). It is also clear that $C$ is the only partial cover
containing $C_l$ and $C_r$ such that $I_t(C)=\lambda_t$.

These considerations lead to the formula
$$w(\lambda_t)=\sum_{(\lambda_l,\lambda_r)} w(\lambda_l)w(\lambda_r)w(S_{\lambda_t}).$$
The sum runs over all pairs $(\lambda_l,\lambda_r)$ that are compatible with $\lambda_t$.
The weight $w(\lambda_t)$ can therefore be computed in a constant number of 
arithmetic operations.

Since the height of $T$ is $O(log(n))$ the
above algorithm can be executed on a circuit of height $O(log(n))$ as well.
\qed

\begin{theorem} \label{hamToCircuit}
The hamiltonian of a $n \times n$ matrix $M$ of bounded treewidth $k$ can be
expressed as a formula of size $O(n^{O(1)})$.
\end{theorem}
\proof
The proof is very similar to that of Theorem~\ref{permToCircuit}.
The only difference is that we only consider partial cycle covers
consisting exclusively of paths,
and at the root of $T$ the partial cycle covers
of the two children must be combined into a hamiltonian cycle.
\qed

\section{Perfect matchings of planar graphs}

In this section we work with undirected graphs.
\begin{definition}
A {\em perfect matching} of a graph $G = \langle V,E \rangle$
is a subset $E'$ of $E$ such
that every vertex in $V$ is incident to {\em exactly one} edge in $E'$.
The {\em weight} of a perfect matching $E'$ is the product of weights
of all edges in $E'$.
By $SPM(G)$ we denote the sum of weights of all perfect matchings of $G$.
\end{definition}

In 1967 Kasteleyn showed in \cite{Ka} that $SPM(G)$ can be computed
efficiently if $G$ is planar. His observations was that for planar graphs
$SPM(G)$ could be expressed as a special kind of Pfaffian. For general
non-planar graphs computing $SPM(G)$ is
$\vnp$-complete (or $\sharpp$-complete in the boolean model of computation).
This follows from the fact that for a bipartite graph $G=(U,V,E)$, $SPM(G)$
is equal to the permanent of its ``bipartite adjacency matrix'' $M$ 
(if $|U|=|V|=n$, $M$ is a $n \times n$ matrix and $m_{ij}$ is equal 
to the weight of the edge between $u_i$ and~$v_j$).
\begin{theorem} \label{planarchar}
Let $(f_n)$ be a family of polynomials with coefficients in a field $K$.
The three following properties are equivalent:
\begin{itemize}
\item $(f_n)$ can be computed by a family of polynomial size weakly
skew circuits.
\item  $(f_n)$ can be computed by a family of polynomial size
skew circuits.
\item There exists a family $(G_n)$ of polynomial size planar graphs
with edges weighted by constants from $K$ or variables of $f_n$ such that 
$f_n=SPM(G_n)$.
\end{itemize}
\end{theorem}
The equivalence of (i) and (ii) is etablished in~\cite{MP} and~\cite{To}.
In \cite{MP} the complexity class $\vdet$ is defined as the class 
of polynomial families computed by polynomial size (weakly) skew circuits,
and it is shown that the determinant is $\vdet$-complete.
We have therefore shown that computing $SPM(G)$ for a planar graph $G$ 
is equivalent to computing the determinant. Previously it was known
that $SPM(G)$ could be reduced to computing Pfaffians~\cite{Ka}.
The equivalence of (iii) with (i) and (ii) follows immediately from 
Theorem~\ref{circuitToSpm} and Theorem~\ref{spmToCircuit}.
\begin{figure}
\begin{center}
\includegraphics[scale=0.4]{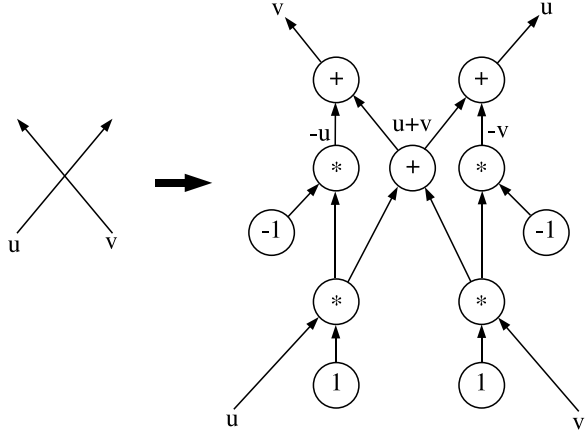}
\caption{Planar crossover widget for skew circuits}\label{planarCrossover}
\end{center}
\end{figure}

\begin{theorem}\label{circuitToSpm}
The output of 
every skew circuit of size $n$ can be expressed as $SPM(G)$
where $G$ is a weighted, planar, bipartite graph with $O(n^2)$ vertices.
The weight of each edge of $G$ is equal to 1, to -1, 
or to an input variable of
the circuit.
\end{theorem}
\proof
Let $\varphi$ be a skew circuit; that is, for each multiplication
gate {\em at least} one of the inputs is an input gate of $\varphi$
(w.l.o.g. we assume it is {\em exactly} one).
Furthermore, by making at most a linear amount of duplication we can
assume all input gates have outdegree 1. Thus, every input gate of $\varphi$
is either input to exactly one addition gate or input to exactly one
multiplication gate (ignoring the trivial special case where $\varphi$ only
consist of a single gate), and throughout the proof we will distinguish
between these two types of input gates.

Consider a drawing of $\varphi$ in which all input gates which are input to
an addition gate, are placed on a straight line, and all other gates
are drawn on the same side of that line. Assume all arrows in the circuit are
drawn as straight lines. This implies at most a quadratic number of places
where two arrows cross each other in the plane. By using the planar
crossover widget from Figure~\ref{planarCrossover} we replace these crossings
by planar subgraphs, introducing at most a quadratic amount of extra gates.

For each multiplication gate we have that exactly one of the input gates
is an input gate of $\varphi$, so these input gates can be placed arbitrarily
close the the multiplication gate in which they are used. Thus we obtain
a {\em planar} skew circuit $\varphi'$ computing the same value as $\varphi$.
\begin{figure}
\begin{center}
\includegraphics[scale=0.8]{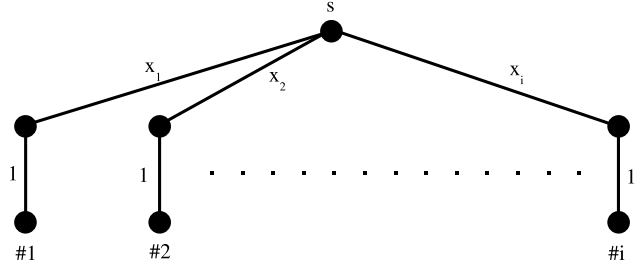}
\caption{Initialization for input gates which are input to addition gates}
\label{planarInit}
\end{center}
\end{figure}

Consider a topological ordering of the gates in $\varphi'$
in which input gates that are input to multiplication gates have numbers
less than 1, and input gates that are input to addition gates have the numbers
1 through $i$ (where $i$ is the number of input gates that are input
to addition gates).
Let $m$ be the number of the output gate in this topological ordering of
$\varphi'$. Steps 1 through $i$ in the
construction of $G$ are shown in Figure~\ref{planarInit}.
Edge weight $x_i$ denote
the input at the gate with topological number $i$ in $\varphi'$.
\begin{figure}
\begin{center}
\includegraphics[scale=0.8]{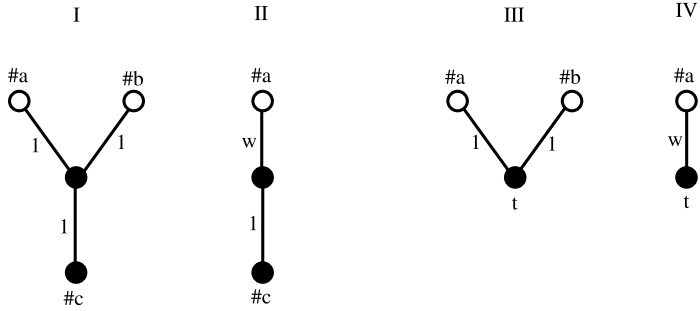}
\caption{I) Non-output addition II) Non-output multiplication
III) Output add. IV) Output mult.}\label{planarIterate}
\end{center}
\end{figure}

For each step $i < m' \leq m$ an addition or multiplication
gate are handled as shown in
Figure~\ref{planarIterate}. White vertices indicate vertices that are
already present in the graph, whereas black vertices indicate new vertices
that are introduced during that step. For multiplications the edge weight $w$
denote the value of the input gate of $\varphi'$, which is input to that
multiplication gate.
Finally, the output gate of $\varphi'$ is handled in a special way.

Correctness can be shown by induction using the following observation.
For each step $1 \leq m' < m$ in the construction of $G$ the following
properties will hold for the graph generated so far:
The labels $\sharp 1, \sharp 2, \ldots ,
\sharp m'$ have been assigned to $m'$ distinct vertices.
For all $1 \leq j \leq m'$ if the vertex with
label $\sharp j$ is removed (along with all adjacent edges), then $SPM$ of
the remaining graph equals the value computed at gate with topological
number $j$ in $\varphi'$.
It is clear that the graph produced during the initialization
(Figure~\ref{planarInit}) has this property.
For the remaining vertices in the topological ordering we either have to
simulate an addition gate ($\sharp c = \sharp a + \sharp b$) or a
multiplication gate ($\sharp c = \sharp a \cdot w$). For each new labeled
vertex added in this way we can see that it simulates the corresponding
gate correctly, without affecting the simulation done by other labeled
vertices in the graph.

Bipartiteness of $G$ can be shown by putting the vertex labeled $s$ as well
as vertices labeled $\sharp i$, $1 \leq i \leq m$,
on one side of the partition, and all
other vertices on the other side of the partition.
\qed

\begin{remark}
The theorem can be proven for {\em weakly} skew circuits as well
without the result from \cite{To} stating that
weakly skew circuits are equivalent to skew circuits. Consider the
graph $G \setminus \lbrace s,t \rbrace$. One can show that this graph
has a single perfect matching of weight $1$.
For simulation of a multiplication gate,
instead of adding a single edge of weight $w$, one can add an entire
subcircuit constructed in the above manner.
\end{remark}

\begin{theorem}\label{spmToCircuit}
For any weighted, planar graph $G$ with $n$ vertices, $SPM(G)$
can be expressed as the output of a skew circuit of size $O(n^{O(1)})$.
Inputs to the skew circuit are either constants or weights of the edges of $G$.
\end{theorem}
\proof
The result will be established by computation of Pfaffians and
is shown by combining results from \cite{Ka} and \cite{MSV}.

Let $H$ be a weighted graph and $\overrightarrow{H}$ an oriented
version of $H$. Then the Pfaffian is defined as:
$$Pf(\overrightarrow{H}) = \sum_{\cal M} sgn({\cal M}) \; w({\cal M}),$$
where $\cal M$ ranges over all perfect matchings of $\overrightarrow{H}$.
The Pfaffian depends on how the edges of $\overrightarrow{H}$ are oriented,
since the sign of a perfect matching depends on this orientation
(details on how the sign depends on the orientation are not needed
for this proof).

It is known from Kasteleyn's work~\cite{Ka} that all planar graphs
have a {\em Pfaffian orientation} of the edges (and that such an orientation
can be found in polynomial time). A Pfaffian orientation
is an orientation of the edges such that each term in the above sum
has positive sign $sgn({\cal M})$.
So for planar graphs computing $SPM(G)$ reduces
to computing Pfaffians (which can be done in polynomial time).

A Pfaffian orientation of $G$ does not depend on the weights of the edges,
it only depends on the planar layout of $G$.
In our reduction to a skew circuit we can therefore assume that a Pfaffian
orientation $\overrightarrow{G}$ is given along with $G$,
thus the problem of computing $SPM(G)$ by a skew circuit
is reduced to computing $Pf(\overrightarrow{G})$ by a skew circuit.

>From Theorem 12 in \cite{MSV} we have that 
$Pf(\overrightarrow{G})$ can be expressed as
$SW(G')$ where $G'$ is a weighted, acyclic, directed graph with
distinguished source and sink vertices denoted $s$ and $t$
(recall $SW(G')$ from Theorem \ref{formulaToPerm}).
The size of $G'$ is polynomial in the size of $\overrightarrow{G}$.

The last step is to reduce $G'$ to a polynomial size skew circuit
representing the same polynomial. Consider a topological ordering of the
vertices of $G'$.
The vertex $s$ is replaced by an input gate with value 1.
For a vertex $v$ of indegree 1,
assume $u$ is the vertex such that there is a directed edge from $u$
to $v$ in $G'$, and assume the weight of this edge is $w$.
We then replace $v$ by a multiplication gate, where one arrow leading
to this gate comes from the subcircuit representing $u$, and the other arrow
leading to this gate comes from a new input gate with value $w$.
Vertices of indegree $d>1$ are replaced by a series of $d-1$ addition gates,
adding weights of all paths leading here, similar to what
is done for vertices of indegree 1.

The circuit produced in this way clearly represent the same polynomial,
and it is a skew circuit because for every multiplication gate at least one
of the arrows leading to that gate comes from an input gate.
\qed

\section{Acknowledgements}

This work was done while U.~Flarup was visiting the ENS Lyon during
the spring semester of 2007.
This visit was partially made possible by funding from
Ambassade de France in Denmark,
Service de Coop\'eration et d'Action Culturelle,
Ref.:39/2007-CSU 8.2.1.



\end{document}